\documentclass[11pt]{IEEEtran}
\usepackage{amssymb}
\usepackage{hyperref}
\usepackage{amsmath}
\usepackage{braket}
\usepackage{graphicx}
\usepackage{color}
\usepackage{qcircuit}

\begin{document}

\title{Fractal Properties of Magic State Distillation}
\author{Patrick Rall, \textit{Quantum Information Center, University of Texas at Austin}\\ \today\\ \vspace{-10mm}}
\maketitle

\newcommand{\eps}{\varepsilon}
\begin{abstract}
    Magic state distillation protocols have a complicated non-linear nature. Analysis of protocols is therefore usually restricted to one-parameter families of states, which aids tractability. We show that if we lift this one-parameter restriction and embrace the complexity, distillation exhibits fractal properties. By studying these fractals we demonstrate that some protocols are more effective when not restricted. Low fidelity states that are usually worthless for distillation are now usable, and fewer iterations of the protocols are needed to reach high fidelity. 
\end{abstract}

Quantum error correction codes allow a quantum computer to protect data from decoherence. Fault-tolerant manipulation of encoded data is challenging since no error correction code supports a universal set of transversal operations \cite{ZCC07}. Restricted sets of operations that can be implemented fault-tolerantly, like the Clifford group, can be elevated to universality via access to certain `magic' states \cite{bk04}. Magic state distillation studies the preparation of these states.

Motivated by the restrictions of fault-tolerant quantum computation, we consider a quantum device with the following ideal operations:
\begin{enumerate}
    \item Clifford operations $\{H, S, CNOT\}$,
    \item measurement in the $\{\ket{0}, \ket{1}\}$ basis,
    \item preparation of stabilizer states,
    \item classical randomness, and
    \item preparation of a state $\rho$, a qubit mixed state.
\end{enumerate}

Two questions arise:
\begin{itemize}
    \item Is the device universal for quantum computation: can $\rho$ in principle be converted to a magic state?
    \item How many copies of $\rho$ are needed to prepare a magic state to a target fidelity? 
\end{itemize}

Despite over a decade of study with many successes, neither question has a complete answer. Ideally we would like a resource theory that can calculate the optimal rate of conversion from copies of $\rho$ to a magic state, and a description of the distillation protocol to do so. Such a theory was recently developed \cite{monotone} for the complete set of stabilizer-preserving operations, which is larger than the set listed above and not known to be achievable in fault-tolerant quantum computation. Veitch et al. \cite{resource} developed a resource theory for the above operations that can upper bound the average rate but not determine possibility of distillation.

Without a resource theory, the traditional approach to attack either question has been to construct and analyze specific protocols built with quantum error correction codes. Analysis is done by examining one-parameter families that consist of low-fidelity versions of the $\ket{H}$ and $\ket{T}$ magic states:
\begin{equation}\rho_T(f) = \ket{T}\bra{T}f + \frac{I}{2} (1-f) \label{eq:oneparamT}\end{equation}
    \vspace{-2mm}
\begin{equation}\rho_H(f) = \ket{H}\bra{H}f + \frac{I}{2} (1-f)\label{eq:oneparamH}\end{equation}

Initial successes by Reichardt \cite{R04} \cite{R06} demonstrated that the vast majority of qubit non-stabilizer states can be distilled by giving explicit codes that increase $f$ when $f$ is above a threshold. However, there remain some states with no known effective distillation protocol, and finding codes with low threshold seems to be limited to guesswork. Concerning fast distillation we can do much better than guesswork using error correction codes with transversal non-Clifford \cite{bravhaah} or Clifford \cite{jones} gates. These techniques almost saturate conjectured asymptotic optimality bounds \cite{infinite}.

In this study we explore what is to be gained from analyzing existing protocols outside of these one-parameter families. We find that some protocols exhibit fractal properties. In section~\ref{sec:background} we give background on analysis of distillation protocols in the mathematical language of fractals. In section~\ref{sec:thresholds} we show that the fractal properties of the five qubit and Steane codes change the picture of which states are known to be distillable in principle. In section~\ref{sec:rates} we analyze the protocols proposed by \cite{bravhaah} and \cite{jones} and study the circumstances under which distillation rate can be improved.

\section{Fatou and Julia sets of Distillation \label{sec:background}}

All  distillation schemes with a single output state can be written as follows \cite{cb-struct}:

\begin{enumerate}
    \item Collect $n$ unentangled copies of an initial state $\rho$.
    \item Project $\rho^{\otimes n}$ onto the codespace of an $[[n,1,d]]$ stabilizer code.
    \item Decode the logical qubit to obtain the output $\rho'$.
\end{enumerate}

In \cite{weightenums} we showed that evolution of $\rho$ under distillation can always be expressed in terms of rational functions. If we use the Bloch sphere expansion,
$$\rho(x,y,z) = \frac{I}{2} + \frac{X}{2}x + \frac{Y}{2}y + \frac{Z}{2}z,$$
then for any $[[n,1,d]]$ code there exist multivariate polynomials $W_I(x,y,z)$, $W_X(x,y,z)$, $W_Y(x,y,z)$ and $W_Z(x,y,z)$ such that the expansion of the output state $\rho'$ satisfies:
\begin{equation}(x',y',z') = \left(\frac{W_X}{W_I}, \frac{W_Y}{W_I},  \frac{W_Z}{W_I}\right).\label{eq:rational}\end{equation}

This map is iterated until a sufficiently high-fidelity state is obtained. From this premise it is almost unsurprising that fractal properties emerge, since fractals are commonly constructed via the iteration of rational functions \cite{Beardon}.

Most fractals are constructed in 2D via the iteration of rational functions in a single complex number, rather than three real numbers. To achieve 3D fractals one can construct rational functions of quaternions. This seems promising since qubit hermitian matrices, e.g., density matrices, are isomorphic to quaternions. However, we find that for all codes we studied it is impossible to rephrase (\ref{eq:rational}) as a single univariate rational function of a quaternion. 

Despite this difference from fractal literature \cite{Beardon} a key concept remains useful: the \textit{Fatou} and \textit{Julia} sets. Let us represent density matrices as vectors $\vec r = (x,y,z)$ in a Euclidian metric space and view distillation as a function $\mathcal{D}(\rho(\vec r)) = \rho' = \rho(\vec r')$. We say $\mathcal{D}$ is \textit{equicontinuous} at $\vec r_0$ if for any $m$-repeated distillation $\vec r \to \vec r^{m}{}'$ and for every $\eps > 0$ there exists a $\delta > 0$ such that:

$$|\vec r - \vec r_0| < \delta \implies |\vec r^{m}{}' - \vec r_0^{m}{}'| < \eps$$

Any $\mathcal{D}$ has a maximal open subset of $\mathbb{R}^3$ that is equicontinuous. We call this the Fatou set $F$, and its complement the Julia set $J$.

Equicontinuity implies that both sets $F$ and $J$ are closed under $\mathcal{D}$. $F$ can \textit{usually} be viewed as the regions of states that converge to a particular fixed point, such as a magic state, or a worthless state like a stabilizer state or the maximally mixed state. $J$ is the boundary between these converging sets and can be viewed as a 3D generalization of a distillation \textit{threshold}, the cutoff between distillable and not distillable states.

Next we give some examples of these sets to introduce the visualization techniques we use for the rest of the paper. Fatou sets are easy to visualize using color. Given an input $\vec r$ we calculate $\vec r^{m}{}'$ for some sufficiently large $m$ such that approximate convergence has been reached. `Small' values, e.g. $m < 15$ are sufficient, although the number of resource states required to produce $\rho^{m}{}'$ is $n^m$ assuming ideal postselection (i.e. not small for a real quantum computer). $\vec r^{m}{}'$ is then assigned to an rgb color where:
\begin{equation} (r,g,b) = \left(\frac{x+1}{2},\frac{y+1}{2}, \frac{z+1}{2} \right) \label{eq:colorscheme} \end{equation}

\begin{figure}
    \includegraphics[width=0.5\textwidth]{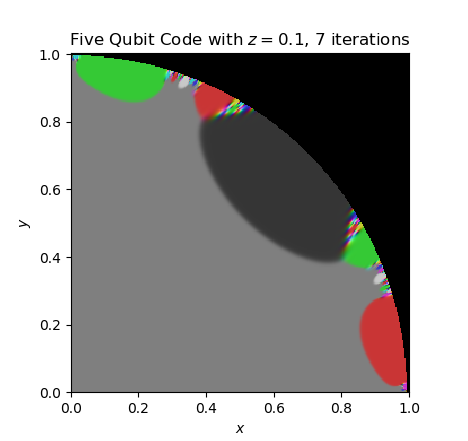}
    \includegraphics[width=0.5\textwidth]{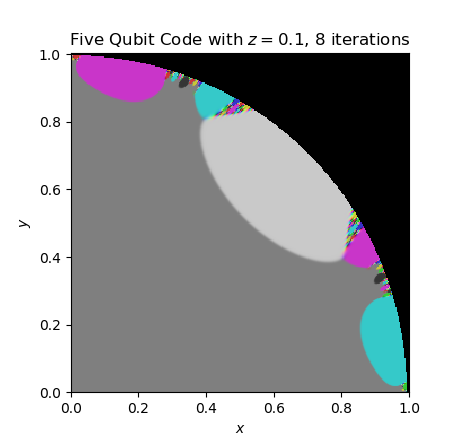}
    \caption{\label{fig:fivequbit-example} A slice of the Bloch sphere showing the fractal properties of the five qubit code (\ref{eq:5qubit}). Position in the diagram encodes the input state, and color encodes the output after several iterations using (\ref{eq:colorscheme}). Table~\ref{tab:tcolor} provides a reference for which colors refer to which states.}
\end{figure}

Any $\vec r$ within a 2x2x2 cube centered at the origin can be represented this way. Fig.~\ref{fig:fivequbit-example} shows the Fatou set structure of the five qubit code, with generators:
\begin{equation}\{XZZXI, IXZZX, XIXZZ, ZXIXZ\}\label{eq:5qubit}\end{equation}

\newcommand\crule[1]{\textcolor[rgb]{#1}{\rule[-0.5em]{1.5em}{1.5em}}}
\begin{table}
    \begin{center}
    \renewcommand{\arraystretch}{1.5}
    \begin{tabular}{ccccc}
        $I/2$  & \crule{0.5,0.5,0.5} & \hspace{2em} & Bad Input & \crule{0,0,0} \\
        $\ket{T}$ & \crule{0.789,0.789,0.789} &&  $S\ket{T}$ & \crule{0.211,0.789,0.789}  \\
        $Z\ket{T}$ & \crule{0.211,0.211,0.789} &&  $S^\dagger\ket{T}$ & \crule{0.789,0.211,0.789}  \\
        $S^\dagger X\ket{T}$  & \crule{0.211,0.211,0.211} & \hspace{2em} & $X\ket{T}$ & \crule{0.789,0.211,0.211}  \\
        $SX\ket{T}$  & \crule{0.789,0.789,0.211} & \hspace{2em} & $ZX\ket{T}$ & \crule{0.211,0.789,0.211}  \\
    \end{tabular}
    \end{center}
    \caption{\label{tab:tcolor}Color encoding of Clifford rotations of $\ket{T}$ using (\ref{eq:colorscheme}).}
    \vspace{-1cm}
\end{table}

\begin{figure}
    \includegraphics[width=0.5\textwidth]{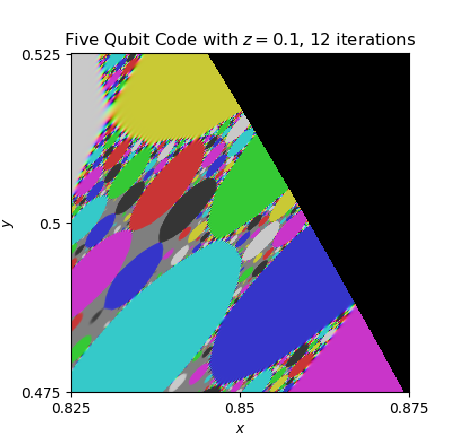}
    \caption{\label{fig:fivequbit-example-zoom} Zoomed version of Fig.~\ref{fig:fivequbit-example}. Detailed regions with self-similarity are visible.}
\end{figure}

    We observe that the majority of states converge to the maximally mixed state, but bubbles near the surface of the Bloch sphere converge to different pure states. The code exhibits cycling behavior: distilling a magic state causes it to rotate by a (possibly state-dependent) Clifford gate. This is a common effect observed previously in both qubit \cite{bk04} and qutrit \cite{qutrit} codes.

    Zooming in on a region in the same diagram we observe fractal properties, as shown by Fig.~\ref{fig:fivequbit-example-zoom}. The structure becomes very detailed in some areas, and exhibits self-similarity. As we render the diagrams in higher detail the edges become blurry. This is because distillation was not iterated infinitely many times and some states have not converged.

The boundaries between the regions of convergence form the Julia set $J$. Points in $J$ are unstable fixed points of distillation up to Clifford rotations. Since $J$ represents the distillation thresholds, we would like a numerical method for identifying points in $J$. 

\begin{figure}[t]
    \includegraphics[width=0.45\textwidth]{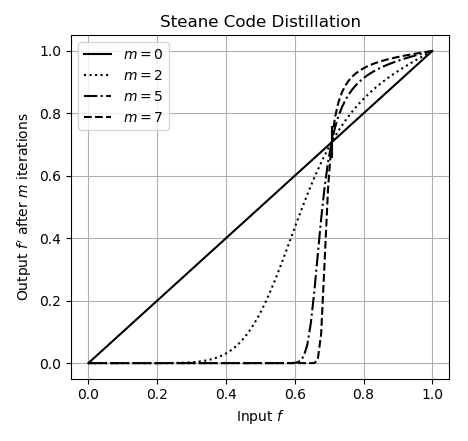}
    \includegraphics[width=0.45\textwidth]{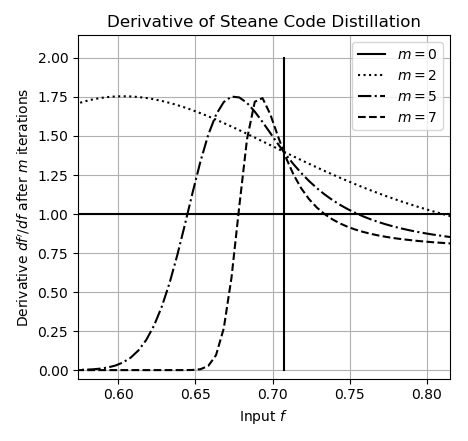}
    \caption{\label{fig:derivative} One-parameter distillation curves of the Steane code for several iterations. We see that the largest derivative is close to the distillation threshold, marked by the black vertical line, for sufficiently large $m$.}
    \vspace{-5mm}
\end{figure}

Points in the Julia set are unstable in the sense that a small perturbation can move them into the Fatou set, and cause them to converge upon distillation. This means a small change in $\vec r$ should cause a large change in $\vec r^m{}'$, which is easily measured using a vector derivative. To illustrate we first consider motion in the one-parameter family $\rho_H(f)$ under the Steane code:

\begin{equation} \label{eq:steane}
\begin{split}
    \{&XXXXIII, XXIIXXI, XIXIXIX,\\ & ZZZZIII, ZZIIZZI, ZIZIZIZ\}.
\end{split}
\end{equation}

Fig.~\ref{fig:derivative} shows the derivative of iterations of the Steane code.  We see that the derivative peaks near the distillation threshold, and that the peak moves closer to the threshold as the code is iterated more often. Thus the derivative serves as a crude numerical means to identify the distillation threshold. In the limit of infinite $m$, the distillation curve should be a step function, and its derivative a $\delta$-function which is zero everywhere except at the threshold.

We generalize this idea to 3D via the directional derivative via a small perturbation $\eps \hat n$. The directional derivative of $m$-repeated application of $\mathcal{D}$, written $\mathcal{D}^m$, in a direction $\hat n$ orthogonal to the Julia set should approach a $\delta$-function. 

$$\lim_{\eps\to 0}\frac{\mathcal{D}^m(\vec r + \eps \hat n) - \mathcal{D}^m(\vec r)}{\eps} = \nabla \mathcal{D}^{m}(\vec r) \cdot \hat n$$

Here $\nabla \mathcal{D}^{m}(\vec r)$ is the Jacobian, and $\hat n$ is a unit vector representing the direction of differentiation. To choose the $\hat n$ such that the derivative is maximized, we take the largest singular value of $\nabla \mathcal{D}^{m}(\vec r)$. This is conveniently encoded by the $L_\infty$ norm: $||\nabla \mathcal{D}^{m}(\vec r)||_\infty$.

\begin{figure}[t]
    \includegraphics[width=0.5\textwidth]{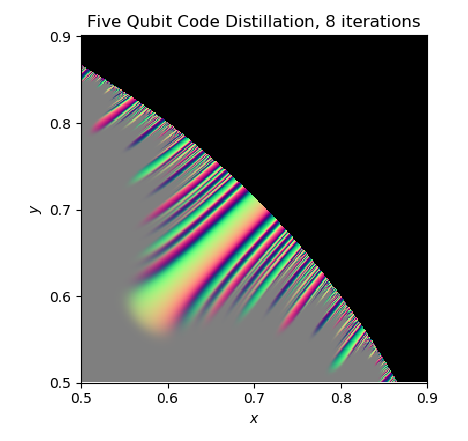}
    \includegraphics[width=0.5\textwidth]{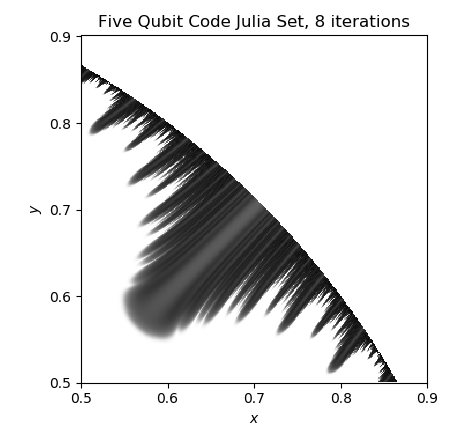}
    \caption{\label{fig:julia-demo} Sketch of the Julia set of the five qubit code using $||\nabla \mathcal{D}^{m}(\vec r)||_\infty$ in the $z= 0$ cross section of the Bloch sphere. Dark regions, indicating high $||\nabla \mathcal{D}^{m}(\vec r)||_\infty$, exhibit fine convergence structure that is very sensitive to the input.}
\end{figure}

We demonstrate the successes and weaknesses of this technique in an interesting region of the five qubit code (\ref{eq:5qubit}) (see Fig.~\ref{fig:julia-demo}). Observe that the large region that converges to the maximally mixed state has small $||\nabla \mathcal{D}^{m}(\vec r)||_\infty$ since it is in the Fatou set. However the region that converges to pure states has oscillation in its output state as a function of input state. Thus $||\nabla \mathcal{D}^{m}(\vec r)||_\infty$ is large in this entire region.

Are these points in the Fatou set or in the Julia set? The uncertainty stems from the fact that only a finite number of iterations was used in the visualization. In fact, this region has fine structure with infinitely many slim bands in the Julia set, separating by the Fatou set. Since a Julia set has no interior \cite{Beardon}, contiguous regions with high $||\nabla \mathcal{D}^{m}(\vec r)||_\infty$ suggest such fine structure. Furthermore, $||\nabla \mathcal{D}^{m}(\vec r)||_\infty$ is only an approximation to the Julia set when $m < \infty$: it may be nonzero outside the Julia set and is only maximal \textit{close} to the Julia set (see Fig.~\ref{fig:derivative}).

In this section we showed how to highlight output states via colors and how to approximate Julia sets via $||\nabla \mathcal{D}^{m}(\vec r)||_\infty$. As an example we demonstrated the fractal properties of the five qubit code. In the next sections we show how analyzing these fractals can benefit distillation protocols.

\clearpage

\section{Distillation thresholds \label{sec:thresholds}}

There are many advantages to restricting to the one-parameter families (1, 2), which is why many studies, e.g., \cite{bk04}, \cite{cb-struct}, \cite{bravhaah}, \cite{jones}, make the simplification. Not only does it make analysis easier, but it is motivated by the computational capabilities of the model. Magic states are eigenstates of Clifford operations:
\begin{equation}
    H\ket{H} = \ket{H}; \hspace{5mm} HS\ket{T} = \ket{T}.
\end{equation}
This allows any qubit state to be projected into the one-parameter families via a `twirling' operation, (where $C$ is some qubit Clifford gate and $0 \leq f \leq 1$):
\begin{equation}
    \rho \to \frac{\rho}{2} + \frac{H\rho H}{2} = C\rho_H(f)C^\dagger
\end{equation} \label{eq:htwirl}
\vspace{-5mm}
\begin{equation}
    \rho \to \frac{\rho}{3} + \frac{HS\rho (HS)^\dagger}{3} + \frac{(HS)^\dagger\rho HS}{3} =C \rho_T(f)C^\dagger
\end{equation}
Twirling is in principle possible in the model by applying a Clifford operation at random. Therefore restricting to one-parameter families is completely feasible.

On the other hand, if Bloch vector length is viewed as a measure of `purity' of a quantum state, the twirling operations actively reduce purity and thus fidelity as a magic state. It is therefore almost surprising that distillation protocols achieve such fast distillation speeds and nearly tight thresholds with this simplification. In this section we illustrate that the five qubit code can only distill certain states if twirling is not used. We also show that twirling provides a limited view of the Steane code that may be an issue when considering noise.

\begin{figure}[h]
    \includegraphics[width=0.5\textwidth]{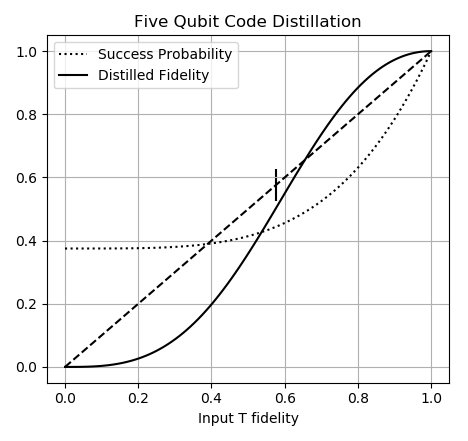}
    \caption{\label{fig:fivequbit-oneparam} Distillation using the five qubit code in the $\rho_T$ family of states. The vertical bar indicates the intersection with the stabilizer polytope. For a small region of states outside the polytope the protocol reduces input fidelity. No code with better $\rho_T$ family performance is known, and it is furthermore impossible for any particular code to achieve tight distillation in the $\rho_T$ family \cite{cb-bound}.} 
\end{figure}

\begin{figure}[h]
    \vspace{-10mm}
    \includegraphics[width=0.5\textwidth]{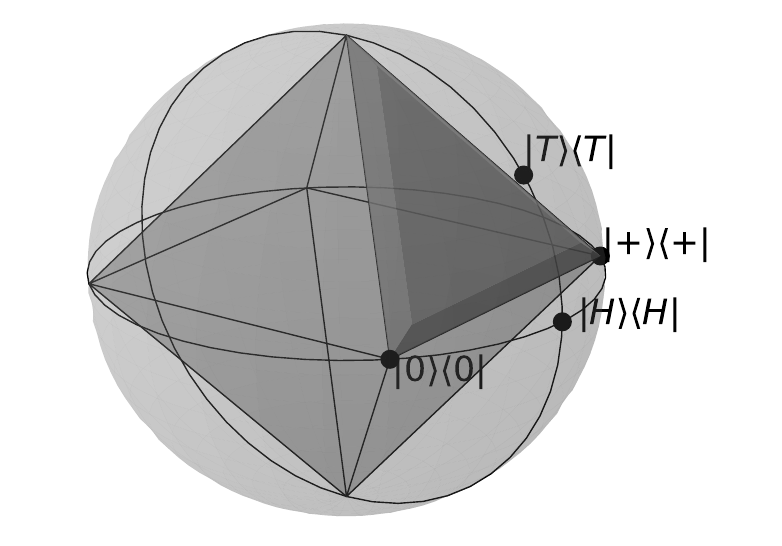}
    \vspace{-8mm}
    \caption{\label{fig:twirled-thresholds} Sketch of distillable and undistillable states using twirling. The polytope of convex combinations of stabilizer states, an octahedron, is marked in medium gray. The Steane code can distill all $\rho_H$ states outside the polytope, and the five qubit code can distill most but not all $\rho_T$ states. The resulting region with no known protocol is highlighted in dark gray. This region is repeated on every face of the octahedron (not shown).
    }
\end{figure}

The five qubit code achieves the best known distillation threshold for distilling states in the $\rho_T$ family. Unlike the Steane code for $\rho_H$ states, the five qubit code is not tight for the $\rho_T$ family. As shown in Fig.~\ref{fig:fivequbit-oneparam} there is a small gap of $\rho_T$ states outside the polytope that converge to the maximally mixed state. This shortcoming manifests itself as a slim polyhedron of states (Fig.~\ref{fig:twirled-thresholds}) for which no successful protocol using twirling is known.

\begin{figure}[h]
    \includegraphics[width=0.5\textwidth]{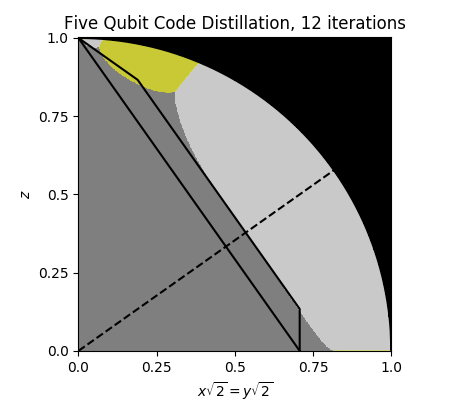}
    \vspace{-8mm}
    \caption{\label{fig:fivequbit-thresholds} Convergence of distillation using the five qubit code. The cross-section intersects the region of states with no known distillation protocol shown in Fig.~\ref{fig:twirled-thresholds}, which is highlighted by the solid lines. A small region of states converging to the state $SX\ket{T}$ (yellow) is clearly distillable to a useful magic state. The dashed line indicates the $\rho_T$ family.}
\end{figure}

However, if we consider a cross-section of the Bloch sphere that intersects this otherwise undistillable polyhedron and plot the convergence under iteration of the five qubit code, we see that there do in fact exist distillable states inside this region (Fig.~\ref{fig:fivequbit-thresholds}). This result is not completely new: it was pointed out by Reichardt \cite{R06} who gave several examples of codes that cut into this region. However it was not realized that the curves describing the boundaries of the region were in fact a part of a fractal. It is therefore probably impossible to describe this boundary analytically.

\clearpage

Our current understanding of low-threshold distillation is unsatisfactory. Resource theories \cite{resource} \cite{monotone} strongly suggest (but do not prove) that every non-stabilizer state should be distillable to a magic state at least in principle. But we have no techniques for constructing codes with low thresholds other than guesswork. The fractal nature of distillation codes may explain why no better techniques are known: thresholds, which are essentially Julia sets, are fundamentally difficult to tailor.

\begin{figure}[t]
    \includegraphics[width=0.5\textwidth]{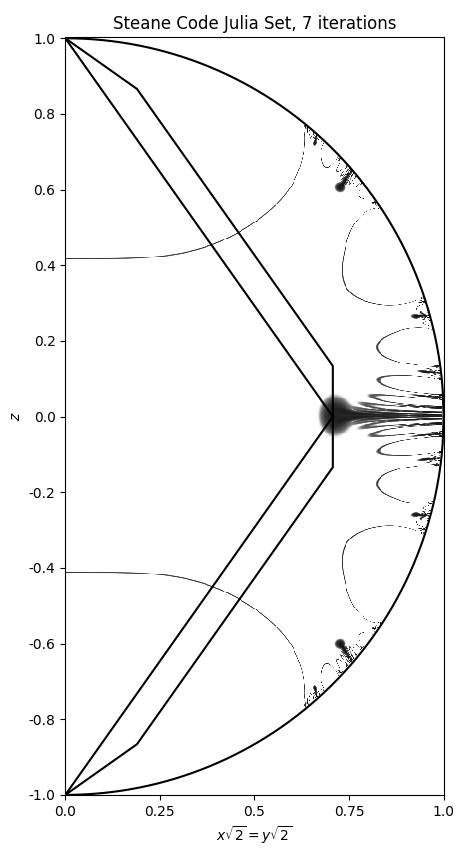}
    \caption{\label{fig:steane-julia} Julia set sketch of the Steane code using $||\nabla \mathcal{D}^{m}(\vec r)||_\infty$. The $\rho_H$ family of states is a horizontal line with $z = 0$, not shown to avoid obscuring the Julia set. The bubbles converge to $\ket{0}$ and $\ket{1}$, whereas the slim dark regions converge to Clifford rotations of $\ket{H}$. We see that the regions that converge to magic states are slim, so a small perturbation could cause a useful state to converge to $\ket{0}$, $\ket{1}$ or $I/2$. The dark region cuts into the stabilizer polytope a little because a finite number of iterations was used (see figure \ref{fig:derivative}).}
\vspace{-10mm}
\end{figure}

Even the simplified picture of best known thresholds using twirling (Fig.~\ref{fig:twirled-thresholds}) is misleading. There exist codes that can penetrate into the region of otherwise undistillable states. But furthermore, the one-parameter simplification provides no direct means for analyzing the stability of protocols to noise other than depolarizing noise (mixing with $I/2$). For example, the Steane code achieves tight distillation of the $\rho_H$ family, thereby distilling the vast majority of non-stabilizer mixed states. If we however plot the Julia set in a surrounding region (Fig.~\ref{fig:steane-julia}) we see that distillation is not stable: small $z$ perturbations can cause distillation to converge to a stabilizer state. This situation is reminiscent of Fig.~\ref{fig:julia-demo}.

In particular, if we simplify the Steane code $W_X, W_Y, W_Z, W_I$ polynomials governing distillation according to eqn. (\ref{eq:rational}) with $\sqrt{2}x = \sqrt{2}y = f$ and assume $z$ is small so we can drop $z^2$ terms and higher powers, we obtain:
$$(f', z') \approx \left(\frac{f^7 + (7/2)f^3}{7f+2}, \frac{-21 f^4 z}{7f+2}   \right)$$

So already at first order Steane code distillation is unstable to perturbations in $z$. The situation for larger deviations is best understood via the numerics in Fig.~\ref{fig:steane-julia}.

Restriction to the $\rho_H$ and $\rho_T$ families is a simplification with many merits. But when studying distillation thresholds, it paints an incomplete picture. Finding distillation protocols for all non-stabilizer states will require analysis of fractals with unpredictable properties.

\clearpage

\section{Distillation rates \label{sec:rates}}

The number of low-fidelity states required to assemble a target state scales exponentially in the number of repeated distillations $m$. For practical distillation protocols it is essential that $m$ is made as small as possible. It was shown in 2009 \cite{cb-struct} that all distillation protocols can be transformed into the standard form described at the beginning of section~\ref{sec:background}. Despite this, protocols designed for distillation with low $m$ are not usually designed in this form.

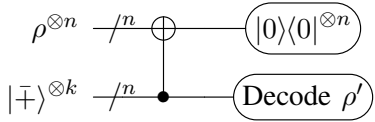
\begin{figure}[h]
    \[
    \Qcircuit @C=1em @R=.7em {
        \lstick{\rho^{\otimes n}} &\qw /^n & \targ & \qw &\measure{\ket{0}\hspace{-1mm}\bra{0}^{\otimes n}} \\
    \lstick{\ket{\bar +}^{\otimes k}} &\qw /^n & \ctrl{-1} &  \qw &  \measure{\mbox{Decode $\rho'$}}
    }
    \]
\caption{\label{fig:bravhaah}Bravyi-Haah magic state distillation using an $[[n,k,d]]$ code. If $\rho= \ket{A}$ the circuit would perform a $T^{\otimes n}$ gate onto the encoded $\ket{+}^{\otimes k}$ state, thereby producing a $\ket{\bar A}^{\otimes k}$ state if the code has a transversal $T$ gate. The input magic states are then projected onto the $\ket{0}$ state and the possibly faulty $\ket{\bar A}^{\otimes k}$ state is projected onto the code space and decoded to yield a $k$ qubit output state $\rho'$. See \cite{bravhaah}.}
\end{figure}

Instead, quantum circuit techniques for assembling magic states using ideal resource state are combined with quantum error correction codes. A state Clifford-equivalent to the $\ket{H}$ state:

\begin{equation}
    \ket{A} = T\ket{+} = \frac{\ket{0} + e^{i\pi/4}\ket{1}}{\sqrt{2}}
\end{equation}

can be used to implement a $T$ gate via a CNOT and a projection onto $\ket{0}$. Using an $[[n,k,d]]$ code where a physical $T^{\otimes n}$ acts like a logical $T^{\otimes k}$, an approximate $T^{\otimes k}$ gate is applied to $\ket{+}^{\otimes k}$ to obtain an approximate $\ket{A}^{\otimes k}$ (figure \ref{fig:bravhaah}). This technique was proposed by Bravyi and Haah in \cite{bravhaah}.

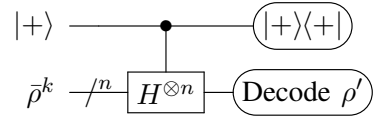
\begin{figure}[h]
    \[
    \Qcircuit @C=1em @R=.7em {
        \lstick{\ket{+}} & \qw & \ctrl{1} & \measure{\ket{+}\hspace{-1mm}\bra{+}}\\
    \lstick {\bar \rho^k}& \qw /^n &  \gate{H^{\otimes n}} & \measure{\mbox{Decode $\rho'$}}
    }
    \]
\caption{\label{fig:codyjones} Cody Jones magic state distillation using an $[[n,k,d]]$ code. $k$ copies of $\rho$ are encoded in a code with a transversal Hadamard gate. An approximate controlled-$H^{\otimes n}$ gates is applied, using more copies of $\rho$ to perform the necessary $T$ gates since controlled-Hadamard is a non-Clifford operation. By projecting onto the $\ket{+}$ state, the encoded $\rho^{\otimes k}$ is effectively projected onto the approximate $+1$ eigenspace of $H^{\otimes k}$ resulting in a possibly higher fidelity $H^{\otimes k}$ eigenstate. See \cite{jones}.}
\end{figure}

Faulty magic states can be used to implement faulty non-Clifford operations, which in turn can be used to implement faulty quantum circuits to improve magic states. This is the principle proposed by Cody Jones in \cite{jones}, detailed in Fig.~\ref{fig:codyjones}: a controlled-Hadamard gate permits projection onto the $+1$ eigenspace of Hadamard. Thereby faulty $\ket{H}^{\otimes k}$ state can be improved using a faulty $\ket{H}^{\otimes 2n}$ state since a controlled-Hadamard requires two $T$ gates. 

Both \cite{bravhaah} and \cite{jones} prepare magic states in Clifford rotations of the $\rho_H$ family. Both papers provide specific quantum codes with the required transversal gate to perform the logical operation that is key to the distillation. \cite{bravhaah} uses `triorthogonal' matrices to construct a family of codes with a transversal $T$ gate, whereas \cite{jones} gives the generators for codes with a transversal Hadamard gate. In principle any code with the required transversal gate and a sufficient minimum distance can be used.

We study three instances of these `fast distillation' protocols:
\begin{enumerate}
    \item ``Bravyi-Haah'': The $\rho^{\otimes 14} \to (\rho')^{\otimes 2}$ protocol based on a [[14,1,2]] triorthogonal code described in \cite{bravhaah}.
    \item ``Cody Jones'': The $\rho^{\otimes 14} \to (\rho')^{\otimes 2}$ protocol based on the [[6,2,2]] code described in \cite{jones}.
    \item ``Cody Jones using Steane Code'': A $\rho^{\otimes 15} \to (\rho')^{\otimes 1}$ protocol applying Fig.~\ref{fig:codyjones} to the [[7,1,3]] Steane code.
\end{enumerate}
The protocols with two output qubits can actually yield a two-qubit entangled state, so $(\rho')^{\otimes 2}$ is misleading. The entanglement becomes negligible for high output fidelity, but cannot be neglected in general. For our purposes we simply discard one of the qubits, since the state is symmetric in the logical bases specified by \cite{bravhaah} and \cite{jones}.

\begin{figure}[h]
    \includegraphics[width=0.5\textwidth]{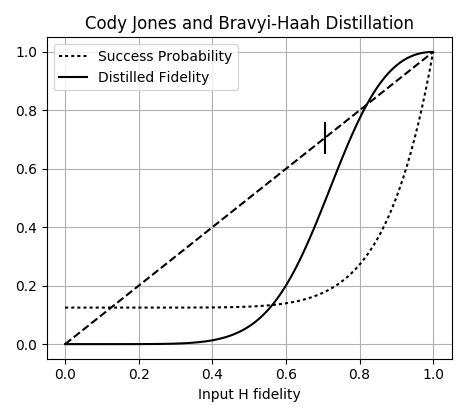}
    \vspace{-10mm}
    \caption{\label{fig:advanced-curve} Distillation curve for Bravyi-Haah distillation, and Cody Jones distillation for both the original code and the Steane code. These three protocols have the same properties in the $\rho_H$ family.}
\end{figure}

\clearpage
\begin{figure}[h]
    (a) 
    \includegraphics[width=0.5\textwidth]{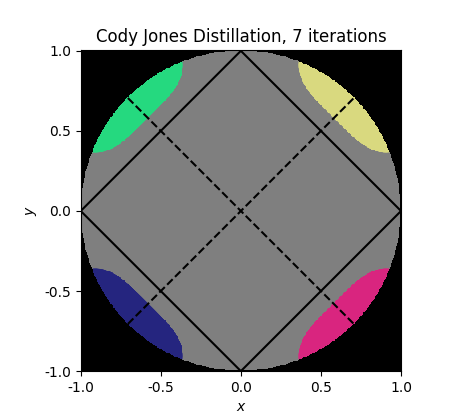}
    (b)
    \includegraphics[width=0.5\textwidth]{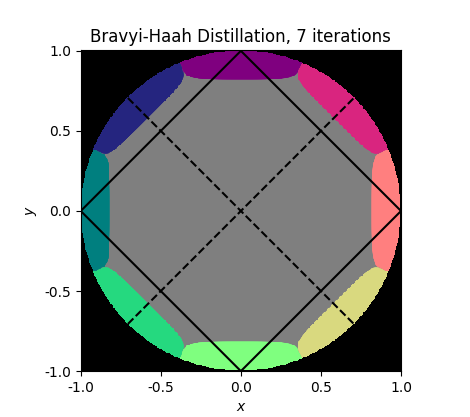}
    (c)
    \includegraphics[width=0.5\textwidth]{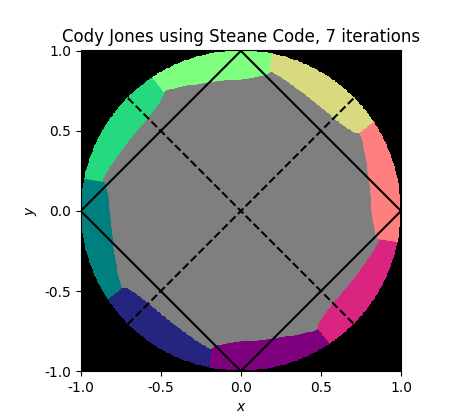}
\end{figure}
\begin{figure}[h]
    (d)
    \includegraphics[width=0.5\textwidth]{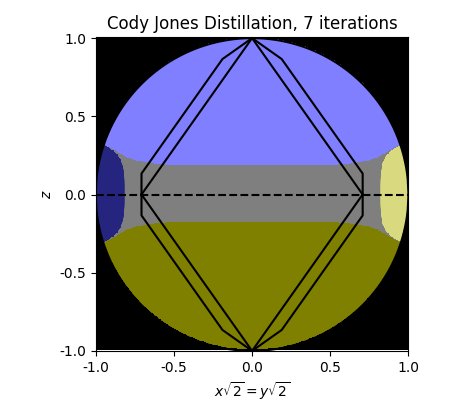}
    (e)
    \includegraphics[width=0.5\textwidth]{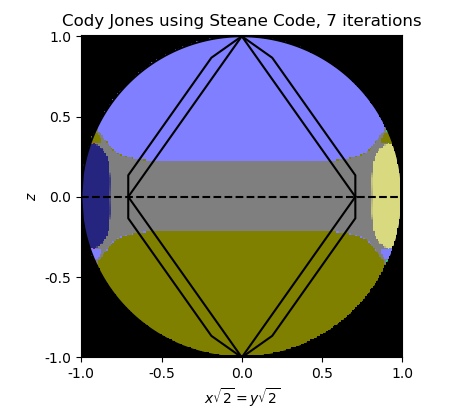}
    (f)
    \includegraphics[width=0.5\textwidth]{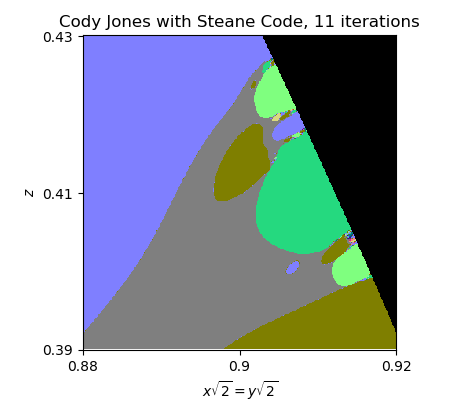}
\end{figure}

\clearpage

\begin{figure}[h]
    \caption{\label{fig:advanced-collage} (Previous page.) Convergence of distillation using the Bravyi-Haah and Cody Jones techniques. (a) (b): in the $xy$-plane, the original versions distill small bubbles near phase rotations of $\ket{A}$. (c): when using the Steane code, the bubbles become asymmetrical. (d): In the $z$, $x$=$y$ plane, original Cody Jones distillation converges to $\ket{0}$ and $\ket{1}$ for many input states. (e), (f): This is still true when using the Steane code, but now fractals appear in some regions.}
\end{figure}

\begin{table}[h]
    \begin{center}
    \renewcommand{\arraystretch}{1.5}
    \begin{tabular}{ccccc}
        $\ket{0}$  & \crule{0.5,0.5,1} & \hspace{2em} & $\ket{1}$ & \crule{0.5,0.5,0} \\
        $\ket{+}$ & \crule{1,0.5,0.5} &&  $\ket{A}$ & \crule{0.854,0.854,0.5}  \\
        $\ket{i}$ & \crule{0.5,1,0.5} &&  $S\ket{A}$ & \crule{0.146,0.854,0.5}  \\
        $\ket{-}$ & \crule{0,0.5,0.5} &&  $Z\ket{A}$ & \crule{0.146,0.146,0.5}  \\
        $\ket{-i}$ & \crule{0.5,0,0.5} &&  $S^{\dagger}\ket{A}$ & \crule{0.854,0.146,0.5}  \\
    \end{tabular}
    \end{center}
    \caption{\label{tab:scolor}Color encoding of the stabilizer states and Clifford rotations of $\ket{A}$ using (\ref{eq:colorscheme}).}
\end{table}

All three protocols behave identically within the $\rho_H$ family (see Fig.~\ref{fig:advanced-curve}), with a non-tight distillation threshold but a fast distillation rate for high initial fidelity: $1-f' \propto (1-f)^2 + O((1-f)^3)$. The analyses within \cite{bravhaah} and \cite{jones} guarantee such good properties provided the codes have a minimum distance $\geq 2$. 

The protocols behave slightly differently outside the $\rho_H$ family, as shown by Fig.~\ref{fig:advanced-collage}. Phase rotations of $\ket{A}$ and in some cases $\ket{+}$ are fixed points of distillation and attract a region around them. We do not observe the detailed self-similarity of fractals in the $xy$-plane. Plotting the Fig.~\ref{fig:codyjones} technique in the $z$, $x$=$y$ plane shows that $\ket{0}$ and $\ket{1}$ attract many states, so input states benefit significantly from being twirled into the $xy$-plane. When using the Steane code with Cody Jones distillation fractals develop, but these do not benefit distillation.

It seems that simple protocols based on Fig.~\ref{fig:bravhaah} and Fig.~\ref{fig:codyjones} do not contain fractals that directly benefit distillation. But this does not mean that it is always better to twirl the input state into the $\rho_H$ family before applying the protocol. To directly compare the performance we repeatedly apply distillation to various input states until a target fidelity is reached.

We compare $M_\text{twirl}$ and $M_\text{no-twirl}$ where, as a function of input state, $M_\text{X}$ is the minimum number of iterations $m$ needed to reach $\rho_H$ fidelity $f > 0.99$. $M_\text{twirl}$ applies a $\rho_H$ twirl (\ref{eq:htwirl}) after every iteration. $M_\text{no-twirl}$ applies a $\rho_H$ twirl only at the end of the distillation so that $\rho_H$ fidelity is well-defined. If $M_\text{twirl} > M_\text{no-twirl}$ then twirling causes distillation to be slower.

\begin{figure}[h]
    \includegraphics[width=0.5\textwidth]{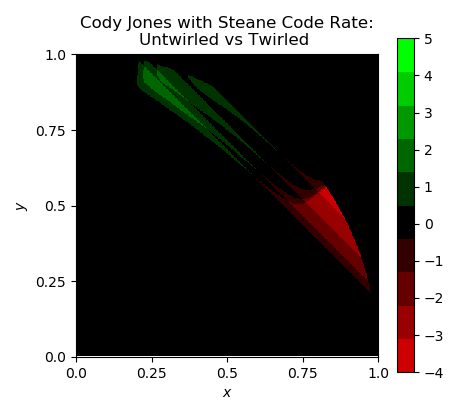}
    \includegraphics[width=0.5\textwidth]{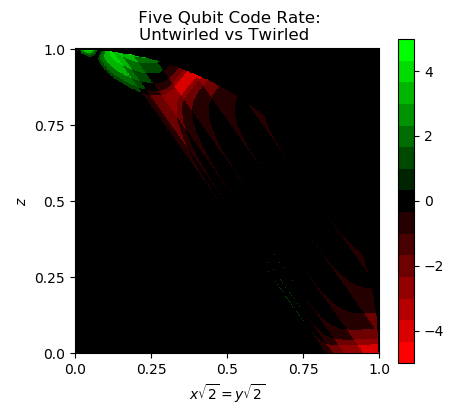}
    \caption{\label{fig:speed} Comparison of the number of distillations required to reach $f > 0.99$ after twirling, with no twirling except at the end ($M_\text{no-twirl}$) or at every iteration $(M_\text{twirl})$ of distillation. $M_\text{twirl} - M_\text{no-twirl}$ is shown as a function of the input state. Green regions with $M_\text{twirl} > M_\text{no-twirl}$ indicate inputs for which twirling is counterproductive to fast distillation. Red regions with $M_\text{twirl} < M_\text{no-twirl}$ benefit from twirling. 
    Compare the top image to Fig.~\ref{fig:advanced-collage}(c) and the bottom to Fig.~\ref{fig:fivequbit-thresholds}.}
\end{figure}

Fig.~\ref{fig:speed} compares $M_\text{twirl}$ and $M_\text{no-twirl}$ for Cody Jones technique using the Steane code, where asymmetry in the $xy$-plane causes some regions to benefit from removing twirling. The five qubit code, in addition to exhibiting one of the best known thresholds in the $\rho_T$ family also has excellent distillation performance there. Fig.~\ref{fig:speed} shows that the five qubit code can even quickly distill some states very close to $\ket{0}$ and $\ket{1}$ if twirling is omitted.

Practical distillation protocols can combine several codes with different properties \cite{fourqubit}. For example, a low threshold code (e.g. Steane) can be used to distill low fidelity states, followed by a code with high distillation rate (e.g. Bravy-Haah). For this reason the Fig.~\ref{fig:speed} analysis is somewhat simplistic since only one code is repeatedly applied. However it is still illustrated that without twirling codes can distill some states faster than others, and that this can be leveraged to improve distillation.

Codes designed for fast distillation also sometimes exhibit fractal properties. Asymmetries around the $\rho_H$ and $\rho_T$ axes present in these fractal codes can cause some regions to distill slower if the inputs are twirled. Analysis of fast distillation protocols is usually done in the asymptotic $(1-f) \ll 1$ regime of a one-parameter family. But this restriction hides much of the complexity of the protocols, which is worth considering since it can give a distillation benefit for low fidelity inputs.

\section{Conclusion}

Quantum computers relying on quantum error correction will likely require magic state distillation as a key element of the architecture. Efficient magic state distillation is thus an important practical goal. 

If analysis of protocols is to be restricted to only certain families of states, the reason should not be just for mathematical convenience. This simplification hides the complex fractal properties that some protocol exhibit, and can paint a misleading picture. There are some input states where codes appear useless or slow when considering only these one-parameter states, when actually the states are useful if twirling is not performed.

Magic state distillation is about crafting fractals that serve our needs. Finding effective protocols is so challenging because under the hood there are mathematical structures with unmatched complexity.

\section{Acknowledgements}

This work was supported by a Research Assistantship at the University of Texas at Austin Computer Science Department, under Dr. Scott Aaronson.


\vfill

\begin{thebibliography}{2}
    \bibitem{ZCC07} Bei Zeng, Andrew Cross, Isaac Chuang. ``Transversality versus Universality for Additive Quantum Codes'' \textit{IEEE Transactions on Information Theory 57 (9), 6272-6284}, \href{https://arxiv.org/abs/0706.1382}{quant-ph/0706.1382} (2007)
    \bibitem{bk04} Sergei Bravyi, Alexei Kitaev. ``Universal Quantum Computation with ideal Clifford gates and noisy ancillas'' \href{http://journals.aps.org/pra/abstract/10.1103/PhysRevA.71.022316}{Phys. Rev. A 71, 022316}, \href{https://arxiv.org/abs/quant-ph/0403025}{quant-ph/0403025} (2004)
    \bibitem{cb-struct} Earl Campbell, Dan Browne. ``On the Structure of Protocols for Magic State Distillation'' \href{https://arxiv.org/abs/0908.0838}{quant-ph/0908.0838} (2009)
    \bibitem{cb-bound} Earl Campbell and Dan Browne. ``Bound States for Magic State Distillation in Fault-Tolerant Quantum Computation'' \href{http://journals.aps.org/prl/abstract/10.1103/PhysRevLett.104.030503}{Physical Review Letters 104 030503} \href{https://arxiv.org/abs/0908.0836}{quant-ph/0908.0836} (2009)
    \bibitem{weightenums} Patrick Rall. ``Signed quantum weight enumerators characterize qubit magic state distillation'' \href{https://arxiv.org/abs/1702.06990}{quant-ph/2017.06990} (2017)
    \bibitem{R04} Ben Reichardt. ``Quantum Universality from Magic States Distillation Applied to CSS Codes''  \textit{Quantum Inf. Proc. 4(3):251-264, 2005}, \href{https://arxiv.org/abs/quant-ph/0411036}{quant-ph/0411036} (2004)
    \bibitem{R06} Ben Reichardt. ``Quantum universality by state distillation'' \textit{Quantum Inf. Comput. 9:1030-1052}, \href{https://arxiv.org/abs/quant-ph/0608085}{quant-ph/0608085} (2006)
    \bibitem{Beardon} Alan F. Beardon. ``Iteration of Rational Functions'' Graduate Texts in Mathematics. Springer-Verlag New York, Inc. 1991.
        \bibitem{bravhaah} Sergey Bravyi, Jeongwan Haah ``Magic state distillation with low overhead'' \href{http://journals.aps.org/pra/abstract/10.1103/PhysRevA.86.052329}{Phys. Rev. A 86, 052329}, \href{https://arxiv.org/abs/1209.2426}{quant-ph/1209.2426} (2012)
    \bibitem{jones} Cody Jones. ``Multilevel distillation of magic states for quantum computing'' \href{https://journals.aps.org/pra/abstract/10.1103/PhysRevA.87.042305}{10.1103/PhysRevA.87.042305} (2012)
    \bibitem{infinite} Jeongwan Haah, Matthew B. Hastings, D. Poulin, D. Wecker. ``Magic State Distillation with Low Space Overhead and Optimal Asymptotic Input Count''  \href{https://arxiv.org/abs/1703.07847}{quant-ph/1703.07847} (2017)
    \bibitem{fourqubit} Adam Meier, Bryan Eastin, Emanuel Knill. ``Magic-state distillation with the four-qubit code'' \href{https://arxiv.org/abs/1204.4221}{quant-ph/1204/4221} (2012)
    \bibitem{trans-class} Jonas Anderson, Tomas Jochym-O'Connor. ``Classification of transversal gates in qubit stabilizer codes'' \href{https://arxiv.org/abs/1409.8320}{quant-ph/1409.8320} (2014)
    \bibitem{resource} Victor Veitch, S A Hamed Mousavian, Daniel Gottesman and Joseph Emerson. ``The resource theory of stabilizer quantum computation'' \textit{New Journal of Physics, Volume 16, No.1}. (2014)
    \bibitem{monotone} Mehdi Ahmadi, Hoan Bui Dang, Gilad Gour, Barry Sanders. ``Quantification and manipulation of magic states'' \href{https://arxiv.org/abs/1706.03828}{quant-ph/1706.03828} (2017)
    \bibitem{qutrit} Hussain Anwar, Earl T Campbell, Dan Browne ``Qutrit magic state distillation'' \href{http://iopscience.iop.org/article/10.1088/1367-2630/14/6/063006/meta}{New Journal of Physics, Volume 14, June 2012} 
    \bibitem{vcge12} Victor Veitch, Christopher Ferrie, David Gross, Joseph Emerson. ``Negative Quasi-Probability as a Resource for Quantum Computation'' \href{http://iopscience.iop.org/article/10.1088/1367-2630/14/11/113011/meta}{New J. Phys. 15 039502}, \href{https://arxiv.org/abs/1201.1256v4}{quant-ph/1201.1256v4} (2012)
\bibitem{rbdobv15} Robert Raussendorf, Dan Browne, Nicolas Delfosse, Cihan Okay, Juan Bermejo-Vega. ``Contextuality and Wigner function negativity in qubit quantum computation'' \href{http://journals.aps.org/prx/abstract/10.1103/PhysRevX.5.021003}{Phys Rev X 5, 021003}, \href{http://arxiv.org/abs/1511.08506}{quant-ph/1511.08506}, (2015)
 \bibitem{rbdobv16} Juan Bermejo-Vega, Nicolas Delfosse, Dan Browne, Cihan Okay, Robert Raussendorf. ``Contextuality as a resource for qubit quantum computation'' \href{https://arxiv.org/abs/1610.08529}{quant-ph/1610.08529} (2016)
       \bibitem{hwve14} Mark Howard, Joel J. Wallman, Victor Veitch, Joseph Emerson. ``Contextuality supplies the magic for quantum computation'' \href{http://www.nature.com/nature/journal/v510/n7505/full/nature13460.html}{doi:10.1038/nature13460}, \href{https://arxiv.org/abs/1401.4174}{quant-ph/1401.4174} (2014)
\end{thebibliography}
\end{document}